\documentclass[11pt,journal]{IEEEtran}
\usepackage{amsmath,amsfonts}
\usepackage{algorithmic}
\usepackage{algorithm}
\usepackage{array}
\usepackage[caption=false,font=normalsize,labelfont=sf,textfont=sf]{subfig}
\usepackage{textcomp}
\usepackage{stfloats}
\usepackage{multirow}
\usepackage{url}
\usepackage{verbatim}
\usepackage{graphicx}
\usepackage{cite}
\usepackage{tabularx}
\usepackage{makecell}
\begin{document}

\title{Magnetic and Mechanical Analysis of Bi-2212 Rutherford Cable in a Cos-Theta Sub-Scale Dipole Coil}

\author{A. D’Agliano, A. V. Zlobin, I. Novitski, D. Turrioni, E. Barzi,~\IEEEmembership{Senior Member,~IEEE}, S. Donati and V. Giusti
\thanks{This work is supported by Fermi Research Alliance, LLC, under contract No. under contract No. DE-AC02-07CH11359 with the U.S. Department of Energy and U.S. Magnet Development Program.

A. D'Agliano is with the Fermi National Accelerator Laboratory, Batavia, IL, 60510, USA, and also with the University of Pisa, Pisa, 56126, Italy (Corresponding author, e-mail: dagliano@fnal.gov).

E. Barzi, I. Novitski, D. Turrioni and A. V. Zlobin are with the Fermi National Accelerator Laboratory, Batavia, IL, 60510, USA.

S. Donati and V. Giusti are with the University of Pisa, Pisa, 56126, Italy.}

\thanks{Manuscript received October 4, 2024; revised January 14, 2025; Accepted for publication February 5, 2025.}}

\markboth{Superconductor Science and Technology,~Vol.~38, No.~3, February~2025}%
{Shell \MakeLowercase{\textit{et al.}}: A Sample Article Using IEEEtran.cls for IEEE Journals}

\IEEEpubid{0000--0000/00\$00.00~\copyright~2024 IEEE}

\maketitle

\begin{abstract}
The U.S. Magnet Development Program (US-MDP) explores high-field accelerator magnets compatible with operational conditions beyond the limits of Nb$_3$Sn technology. The ongoing R\&D High-Temperature Superconductors (HTS) suggests using Bi$_2$Sr$_2$CaCu$_2$O$_{8-x}$ (Bi-2212) as superconducting element. Bi-2212 Rutherford cables maintain a high critical current (I$_C$) when exposed to a large external magnetic field. However, Bi-2212 exhibits an oversensitive stress-strain response when subject to large Lorentz forces. 
This paper reports on the magnetic and mechanical analysis of the Bi-2212 cosine-theta insert being developed at Fermilab for a hybrid magnet composed of two external layers of Nb$_3$Sn and two internal layers of Bi-2212. 
We performed a FEM analysis of the insert to estimate the HTS stress state in the coil's strands under magnetic and mechanical loads.

\end{abstract}


\section{Introduction}

\IEEEPARstart{I}{n} the past few years, scientists have gained increasing interest in realizing high-field magnets using HTS-conductors. The research team at the Lawrence Berkeley National Laboratory (LBNL) has developed and tested magnet models based on canted-cosine-theta, racetrack, and pancake coils \cite{ref1, ref2}.
Fermilab is developing a cosine-theta dipole insert based on a stress-managed structure and a Bi-2212 Rutherford cable \cite{ref3, ref4}.

Bi-2212 has been chosen since it can carry a large current density (J$_E$) in a high external magnetic field. Bi-2212 could allow breaking the threshold of a 20 T magnetic field inside a dipole bore \cite{20T}, which is one of the main goals of the US Magnet Development Program \cite{USMDP}.
The insert will be placed in the innermost part of a multi-layer hybrid magnet and, thus, exposed to an external magnetic field that could exceed 14.6 T, which is the record value for a cosine-theta dipole bore field achieved at Fermilab during the test of the Nb$_3$Sn Cosine-Theta MDPCT1 in 2020 \cite{rec}. Experimental tests demonstrated that Bi-2212 can still carry a current density above 1000 A/mm$^2$ at 14.6 T \cite{ref5}.
This makes Bi-2212 one of the best candidates for significantly improving high-field magnets' performance for accelerator applications.

\begin{figure}[!ht]
\centering
\includegraphics[width=3.4in]{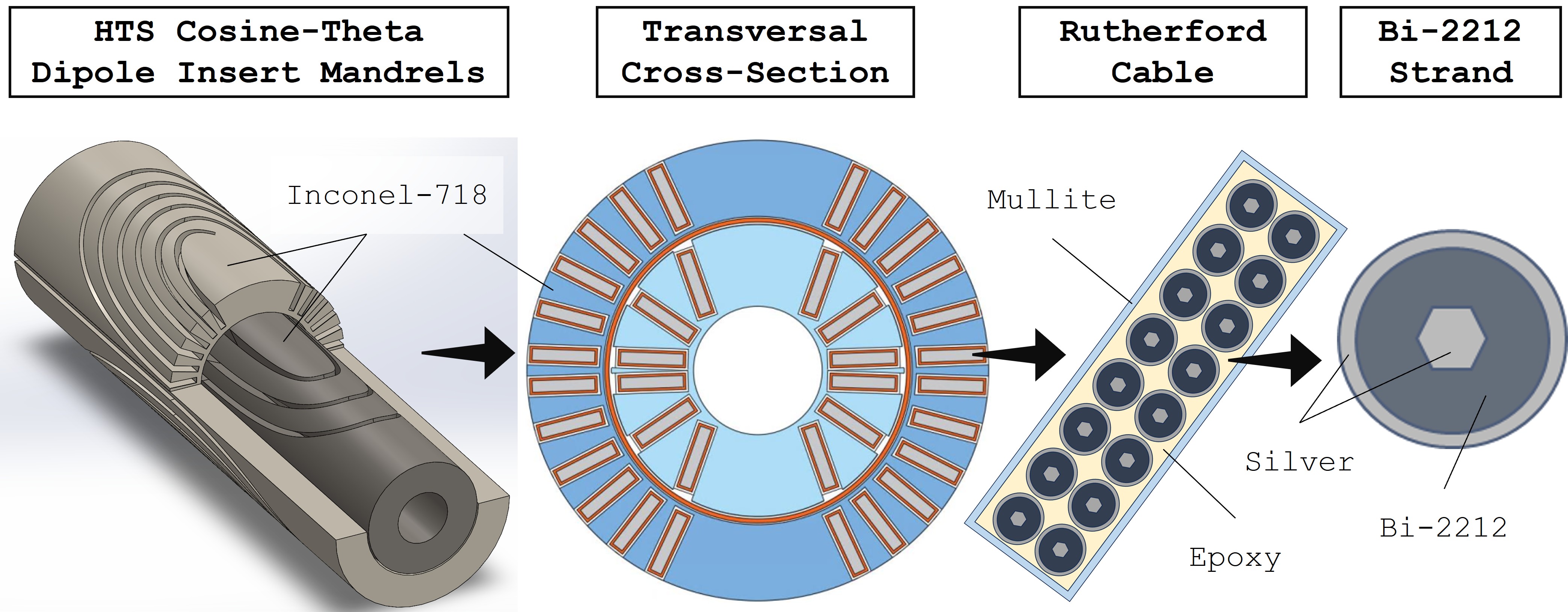}
\caption{Schematic representation of the dipole insert and Bi-2212 Rutherford cable as implemented in the Ansys FEM simulation (heterogeneous model).}
\label{fig:cabdet}
\end{figure}

\IEEEpubidadjcol

Fermilab's team aims to characterize the magnetic and mechanical behavior of a Bi-2212 small-aperture cosine-theta insert (Fig. \ref{fig:cabdet}) with a simulation and validate the readiness of magnet design for fabrication and test.
We used the Ansys Parametric Design Language (APDL) to perform a 2D FEM analysis. In \cite{ref6} has been reported a study of a previous version of the insert for which a simplified model of the cable, defined as a rectangle of homogeneous material placed inside the mandrel grooves, was adopted. This cable model, referred to as the "homogeneous model" in the following, provides an approximate estimate of the equivalent stress in the coil.
To improve the simulation quality, we introduced a more detailed model of the cable to distinguish the relevant parts of the conductor: a) the Ag hexagonal core and outer strand ring, b) the Bi-2212 conductor area, c) the Mullite sleeve as insulation, d) the epoxy in all the gaps (Fig. \ref{fig:cabdet}). This model will be referred to as the "heterogeneous model".
The heterogeneous model allows to specify the pure isotropic material properties inside the cable cross-section (i.e., mechanical characteristics, stress-strain behavior, current degradation related to applied load) and to obtain a more accurate estimate of the stress and strain distributions in the conductor. 
This paper describes the most recent design of the Bi-2212 cosine-theta dipole model being developed at Fermilab. It also reports the results of magnetic and mechanical analyses performed in the two configurations of homogeneous and heterogeneous Bi-2212 cable models.

\section{HTS cosine-theta Bi-2212 Coil Design}

\subsection{Cross-Section Geometry Evolution}

The insert coil was designed to fit into the 60 mm aperture of the outer dipole \cite{NB3SN} realized at Fermilab with two layers of Nb$_3$Sn. The mandrel is the insert's main structural part and provides mechanical support to the coil turns during the winding process, the heat treatment, and the coil powering.
Table \ref{tab:geomelepar} reports the geometrical and electrical parameters of Rutherford cable and Bi-2212 strands \cite{ref4}. The position of the Rutherford cables in the coil transversal and axial cross-sections and the mandrel geometry have been chosen to maximize the magnetic field intensity and the field quality in the bore using ROXIE code \cite{ref7}.

\begin{table}[!ht]
\centering
\caption{Bi-2212 Rutherford Cable and Strand Parameters\label{tab:geomelepar}}
\begin{tabular}{p{4.5cm} c c}  
\hline
 \textbf{Parameter} & \textbf{values} & \textbf{unit} \\
 \hline
 Cable ID & LBNL - 1110 & \\ 
 Number of strands & 17 &  \\
 Bare cable width & 7.8 & mm \\
 Bare cable thickness & 1.44 & mm \\
 Cable transposition pitch & 58 & mm \\ \hline
 Billet ID & PMM180207-2 & \\
 Strand diameter before/after reaction & 0.80/0.778 & mm \\
 Strand architecture & 55 x 18 & \\
 Strand fill factor & 23 & \% \\
 Strand twist pitch & 25 & mm \\
 Strand $I_c$ (4.2K, 5T) & 460-640 & A/mm$^2$ \\
 \hline
\end{tabular}
\end{table}

\begin{figure}[!ht]
\centering
\includegraphics[width=3.4in]{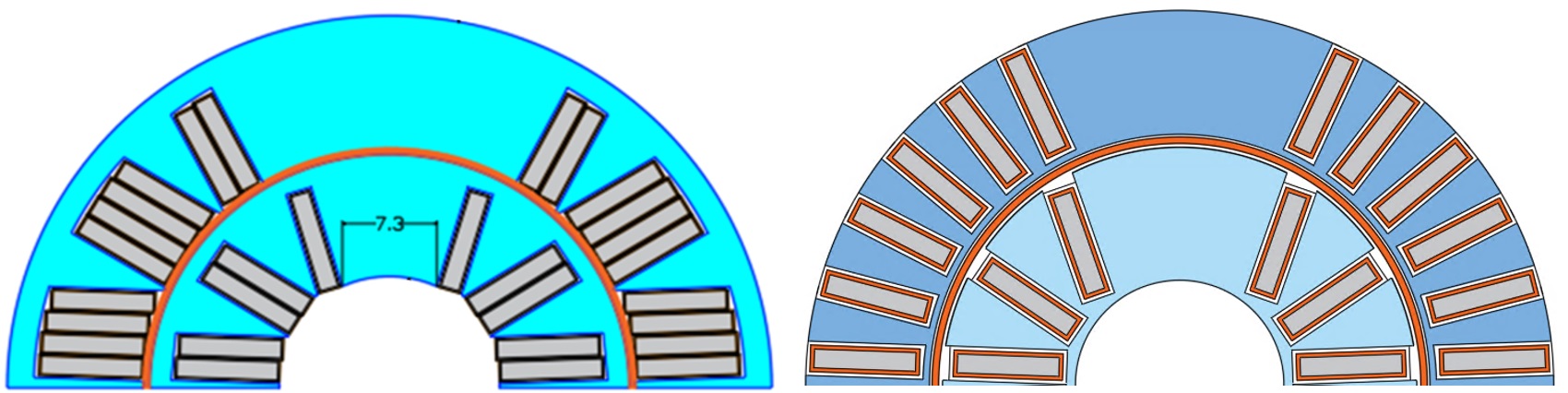}
\caption{Schematic representation of the insert transverse cross-section. The original design is on the left, and the final design is on the right.}
\label{fig:mver}
\end{figure}

The mandrel structure has evolved \cite{ref6} to optimize the magnetic parameters and minimize the bending radius of the turns located at the cosine-theta pole regions.
Fig. \ref{fig:mver} shows the original 2-layer 15-turn insert design as described in \cite{ref6} and the final 2-layer 9-turn insert design analyzed in this paper. The original design allowed the cable to be wound from the cylinder's internal surface to improve mechanical robustness.
However, the minimum bending radius in the pole region was approximately 3.6 mm, which was too small to guarantee the integrity of the Bi-2212 strands.
The final design addressed this issue by reducing the number of turns in both layers 
and increasing the minimal bending radius to approximately 5.5 mm which ensures the integrity of the Bi-2212 strands.
To minimize the risk of damaging the cable and simplify the winding procedure, the turns are inserted in the grooves from the external surface of the barrel, at the slight expense of the coils' radial stiffness.

\begin{figure}[!ht]
\centering
\includegraphics[width=3.2in]{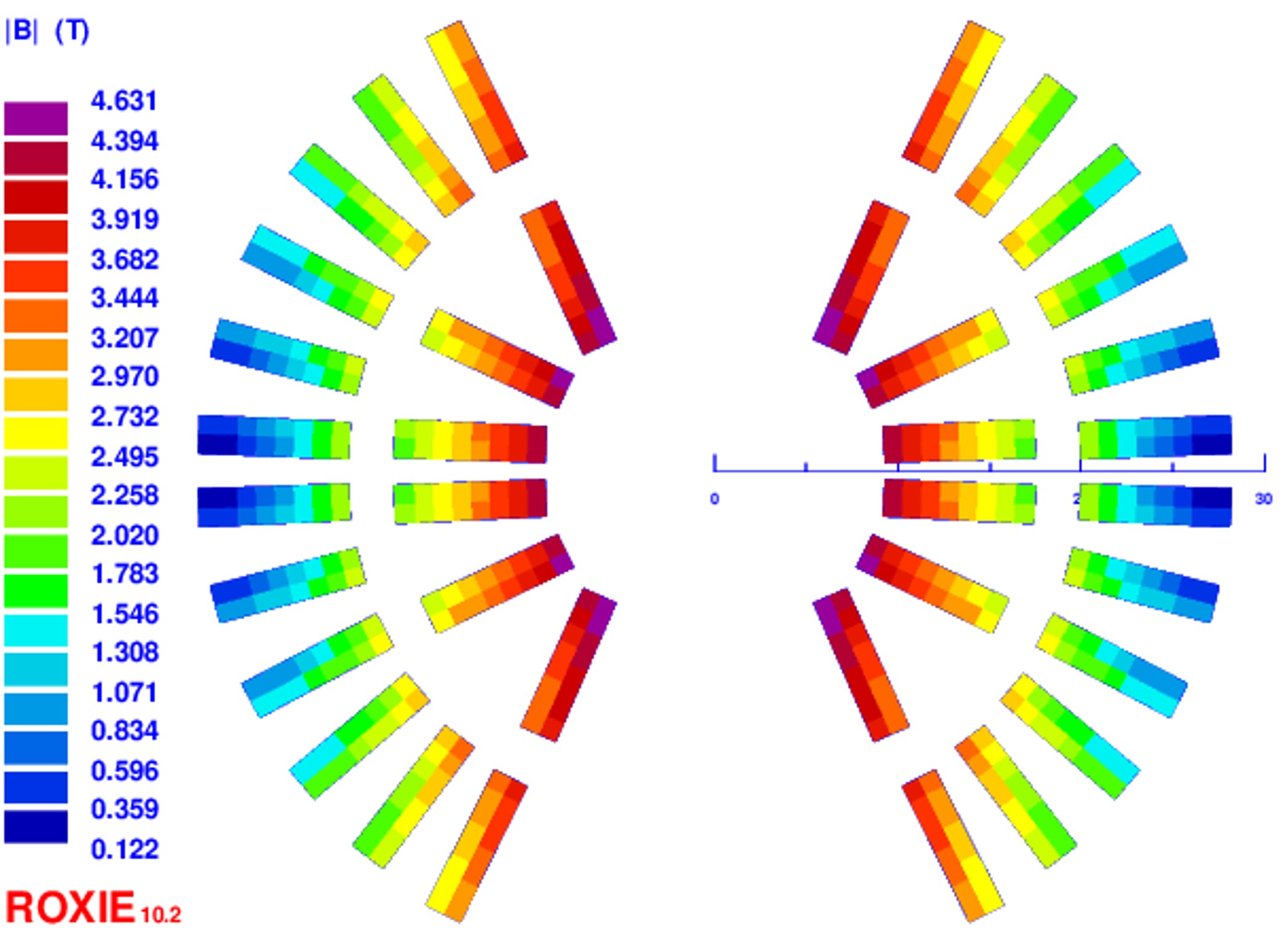}
\caption{Magnetic field intensity for the HTS insert dipole at 10 kA as determined with ROXIE.}
\label{fig:ROXIEmerged}
\end{figure}

Fig. \ref{fig:ROXIEmerged} shows the magnetic diagrams in the coil turns calculated with ROXIE at 10 kA. In the simulation, the final insert design was modeled inside an iron yoke structure characterized by a 60 mm inner diameter and a magnetic relative permeability of 1000.
The area in the aperture where the relative error between the B$_1$ field component and the field computed considering all the harmonics \cite{ref8} is below 10$^{-4}$, has an approximate diameter of 5.4 mm.

\subsection{Insert Coil Practice Winding}

\begin{figure}[!ht]
\centering
\includegraphics[width=3.2in]{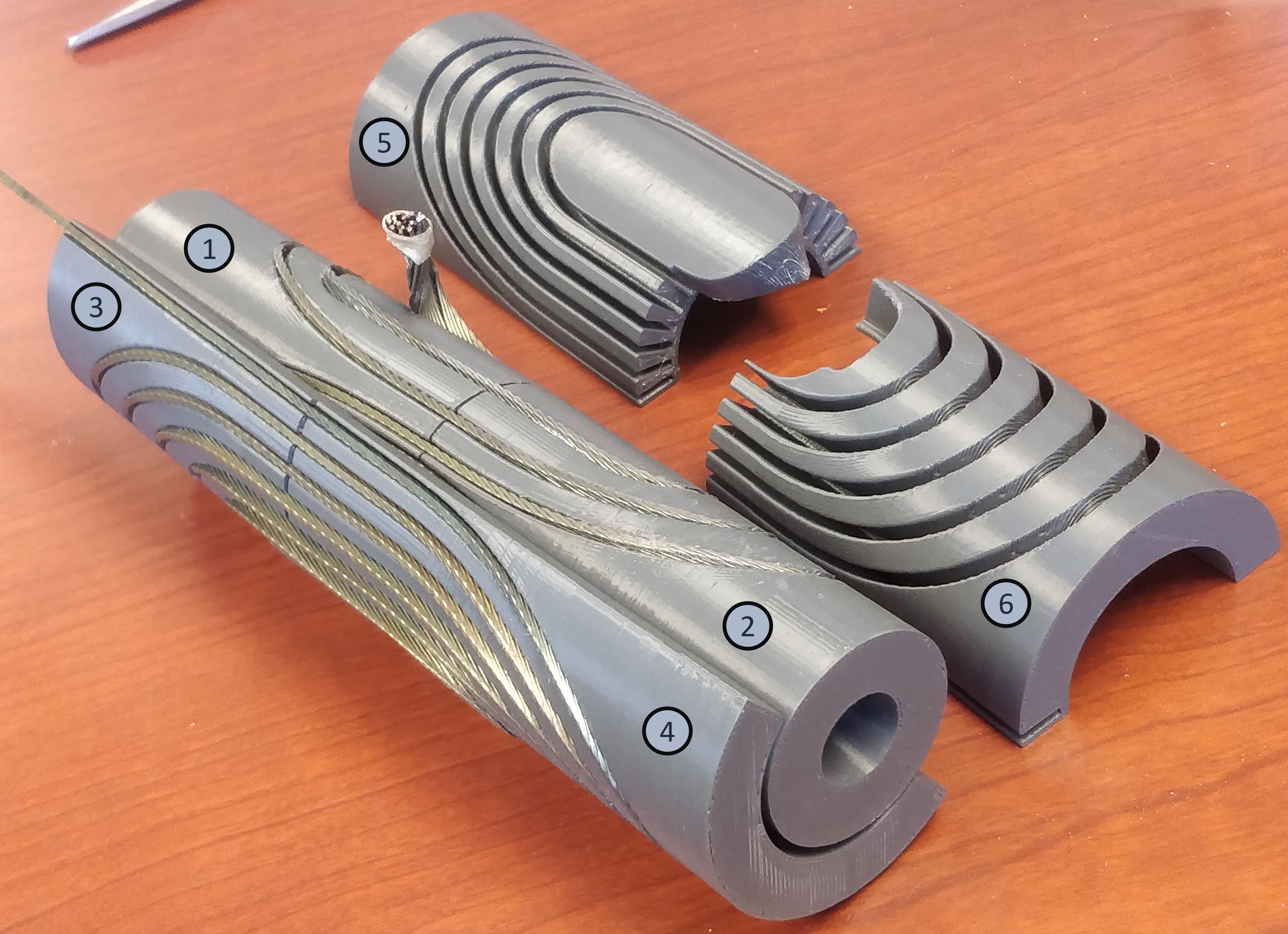}
\caption{3D-printed 2-layer mandrel used to test the winding feasibility and validate the procedure.}
\label{fig:winded}
\end{figure}

Fig. \ref{fig:winded} shows the six mandrel pieces that constitute the mechanical structure of the insert coil. The two inner parts (number 1 and number 2 in the picture) support the inner coil, while the four outer parts (number 3,4,5,6) support the outer coil. The insert internal diameter is 15.5 mm, and the external diameter is 57.6 mm, with a gap of 0.5 mm between the two layers to insert Mullite insulation \cite{ref7}.
Coil winding starts from the interface between parts 1 and 2 (Fig. \ref{fig:winded}). 
The transition from the inner to the outer layer is performed 
with a layer jump at the pole region. The Rutherford cable lead end enters the mandrel from the same side where the return end exits. In this way, the two ends connect with the power supply splices on the same side of the insert. The winding procedure has been successfully tested with a dummy cable.

\section{BI-2212 Insert FEM Analysis}

\subsection{Magnetic Analysis}

We performed the magnetic analysis of the Bi-2212 insert with the homogeneous and the heterogeneous models. In both cases, we used the ANSYS element type \textsc{Plane 53}, which models 2-D planar and axisymmetric magnetic fields. The simulation is planar, and proper boundary conditions guarantee compliance with the symmetry condition. The element is defined by 8 nodes and is based on the magnetic vector potential formulation for the computation of the magneto-static analysis.
\textsc{Plane 53} allows the introduction of a user-defined iron B-H curve. The symmetry constraints are applied on the vertical axis and the outer cylindrical lines of the outer air area to impose that the z-component of the magnetic vector potential is equal to zero. We applied an I$_C$ load of 10 kA distributed on the Bi-2212 strands in the cable (Fig. \ref{fig:cabdet}).
The total current load corresponds to a J$_E$ of approximately 1900 A/mm$^2$ (at 5 T, 4.2 K). This current density is consistent with the data of the Bi-2212 short samples tested at these specific values of the magnetic field and temperature \cite{ref5}.

\begin{table}[!ht]
\centering
\caption{Bi-2212 Insert Parameters at 4.2 K\label{tab:maganalysis}}
\begin{tabular}{p{4.5cm} c c}  
\Xhline{3\arrayrulewidth}
\multirow{2}{*}{\thead{\textbf{}}} &       \multicolumn{2}{c}{\thead{\textbf{HTS Insert}}}     \\
\cline{2-3}
                                            & \thead{\textbf{Inner coil}} & \thead{\textbf{Outer coil}} \\
\hline
Bore field [T] &  \multicolumn{2}{c}{4.36} \\ 
Peak field [T] &  4.85  &  3.95 \\ 
Current [kA] & \multicolumn{2}{c}{10.0} \\ 
Inductance [mH/m] & \multicolumn{2}{c}{0.21} \\ 
Stored Energy @10 kA [kJ/m] & \multicolumn{2}{c}{10.7} \\ 
F$_r$ [MN/m] & 0.07 & 0.04\\ 
F$_a$ [MN/m] & -0.06 & -0.09\\ 
Number of turns & 3 & 6 \\ 
\Xhline{3\arrayrulewidth}
\end{tabular}
\end{table}

\begin{figure}
\centering
\includegraphics[width=3.2in]{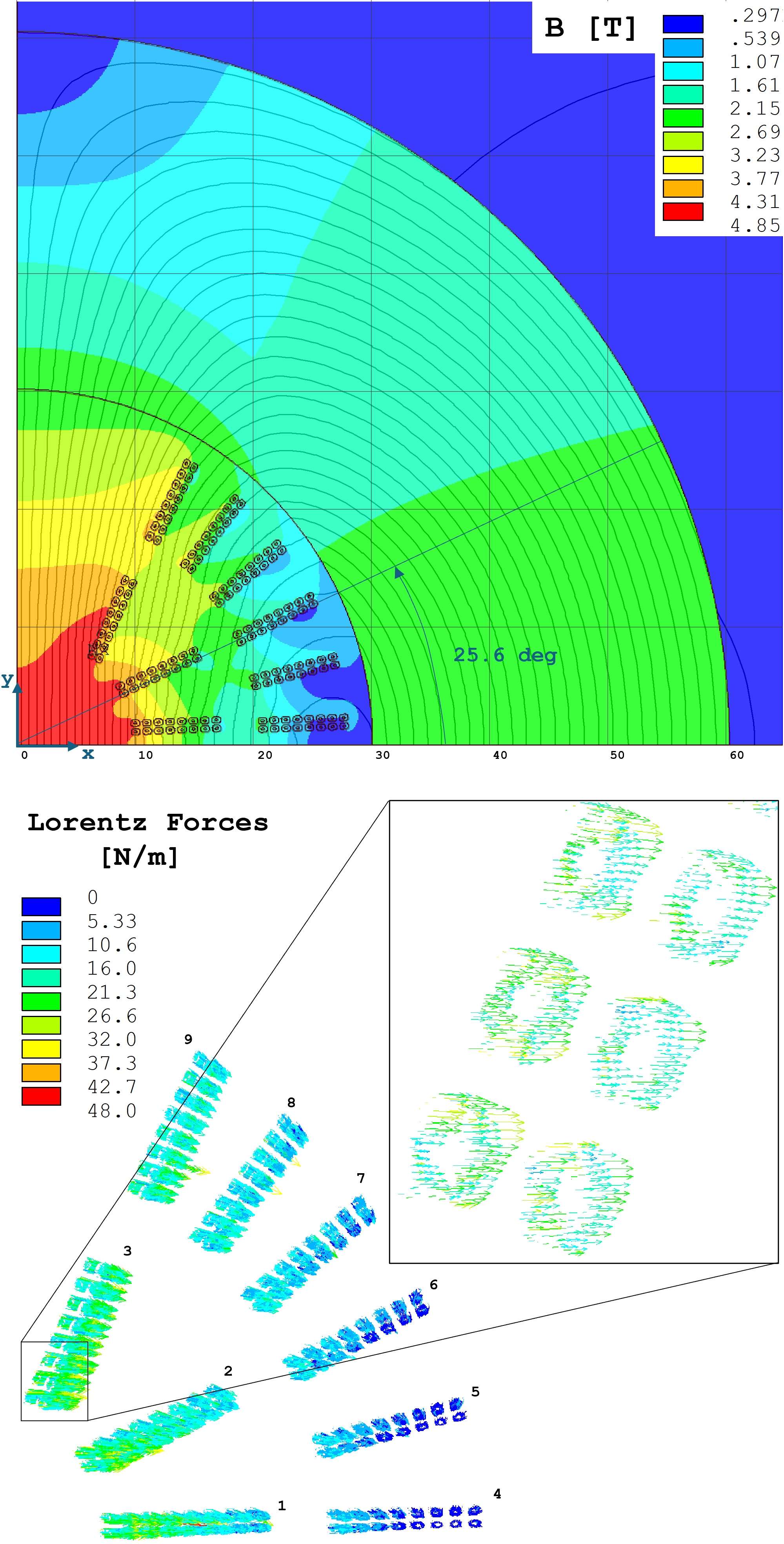}
\caption{Magnetic field intensity in the aperture, coils, and iron yoke (Top) and Lorentz forces in the Bi-2212 strands (Bottom) for the heterogeneous model as determined with ANSYS.}
\label{fig:mag_res}
\end{figure}

We implemented a detailed geometry of the cable components in the heterogeneous model: a) the Bi-2212 region, b) the silver core and outer ring of the strand, c) the Mullite insulation area, and d) all the gaps filled with epoxy. After importing the ROXIE coordinates of each cable in the ANSYS code, a user-defined APDL macro created all the coil turns at the designed locations.
The magnetic simulation of the heterogeneous model shows that with the nine turns of the Bi-2212 cable, a maximum field of 4.85 T can be generated in the coil and 4.36 T in the bore at 4.2 K. The maximum difference between the magnetic field computed with ANSYS in the heterogeneous conductor and ROXIE is of the order of 4.7\% (4.63 T at 10 kA, Fig. \ref{fig:ROXIEmerged}), and it is 2.3\% if the ANSYS homogeneous model is considered. 

Fig. \ref{fig:mag_res} shows the distribution of the magnetic field and the Lorentz force as provided by the ANSYS simulation. Table \ref{tab:maganalysis} summarizes the main magnetic parameters of the Bi-2212 insert. The total Lorentz forces are the sum of all the nodal results on the turns of the specified layer and are reported in a cylindrical coordinate system (radial and azimuthal components).
 
\begin{figure}[!ht]
\centering
\includegraphics[width=3.4in]{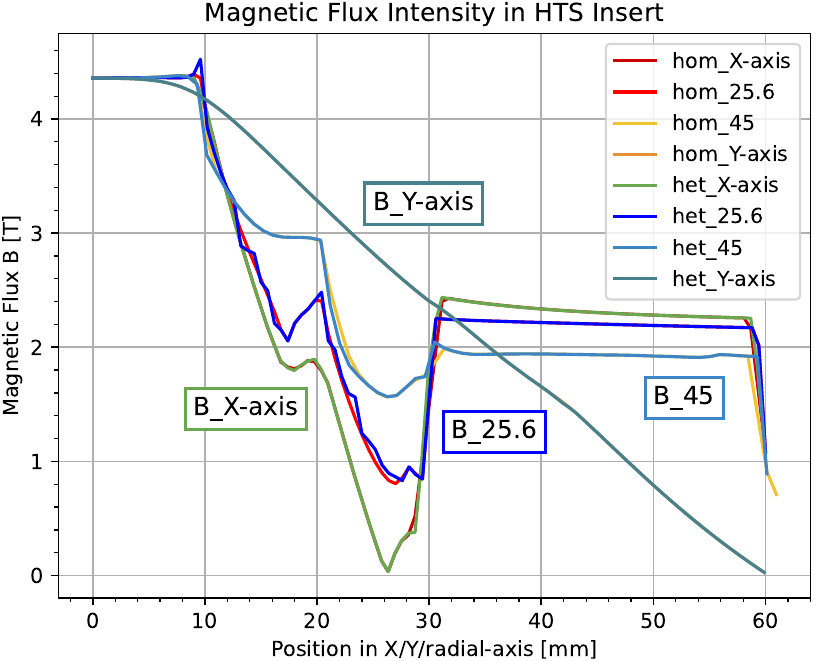}
\caption{Magnetic field intensity for the HTS insert dipole as determined along the $x$-axis, $y$-axis, the line at 25.6 degrees, and the bisector of the first quadrant. 
The '\texttt{hom\_}' and  '\texttt{het\_}' refer to the homogeneous and heterogeneous models, respectively.}
\label{fig:Bpath}
\end{figure}

Fig. \ref{fig:Bpath} shows the magnetic field intensity along specific directions for the homogeneous and the heterogeneous models. The fields computed for the two models are almost identical, except for the positions of the nodes, which define the conductor components in the heterogeneous model. The maximum difference between the two models is 0.18 T along the direction at 25.6 degrees from the $x$-axis, which intersects the conductors of layers 1 and 2 (Fig. \ref{fig:mag_res}).
We verified that the two models provide almost identical nodal sums of the Lorentz forces as computed on each coil.

\subsection{Conductor Mechanical Properties}

\begin{figure}[!ht]
\centering
\includegraphics[width=3.2in]{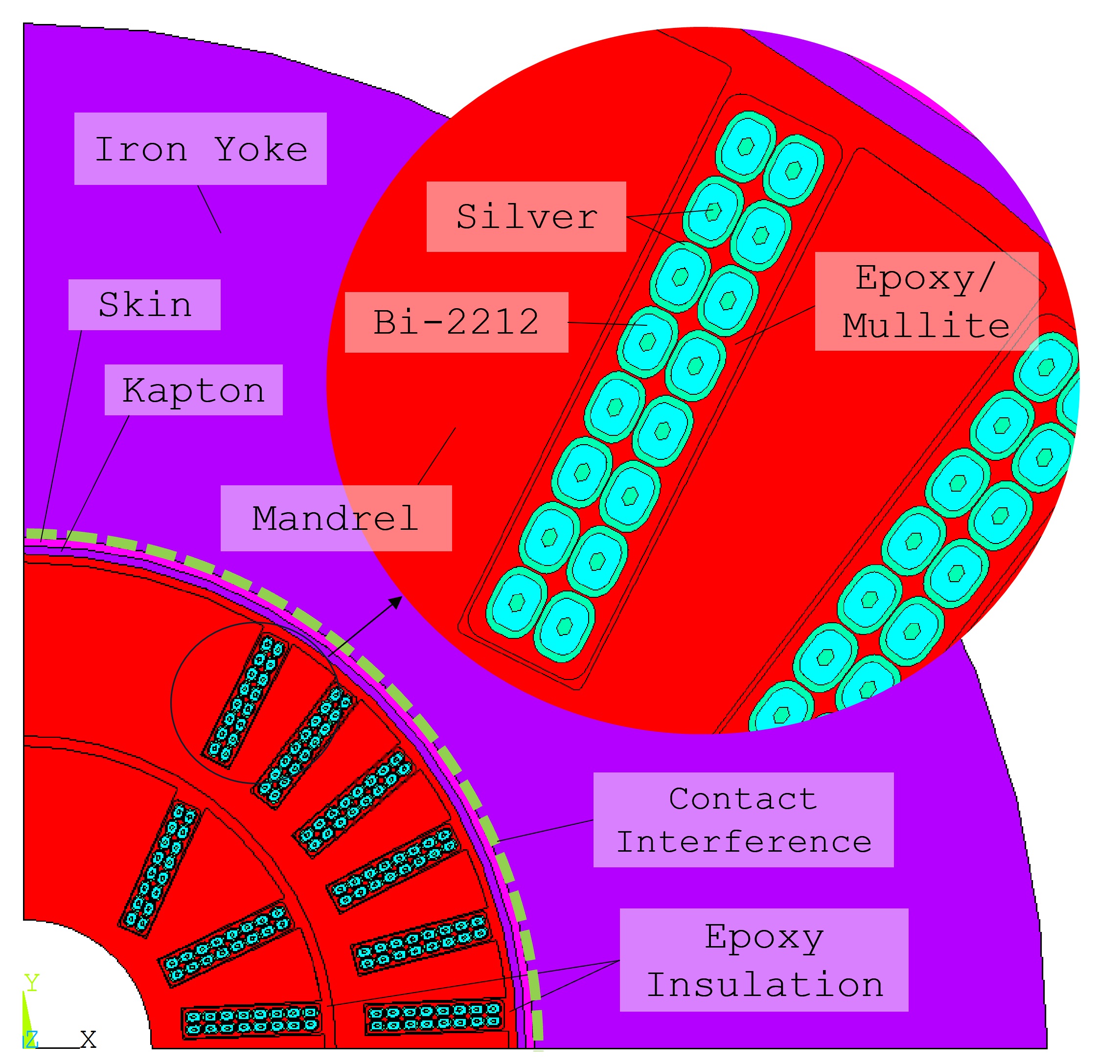}
\caption{Geometry of the Bi-2212 sub-scale dipole insert modeled in ANSYS mechanical analysis with the cable described according to the heterogeneous model. The structural components are also highlighted.}
\label{fig:mech_struct}
\end{figure}

Fig. \ref{fig:mech_struct} shows the geometrical description of the Bi-2212 sub-scale dipole insert as modeled for the ANSYS mechanical analysis according to the heterogeneous model, including all the structural and cable components.
One of the major challenges of a FEM analysis is the definition of the material properties. To better define the mechanical properties of the Bi-2212 strands, we measured stress-strain curves of heat-treated and non-heat-treated Bi-2212 strand samples with a tensile test Instron machine.
Both types of Bi-2212 specimens were 150 mm long with a 0.8 mm diameter. Each strand edge was glued with epoxy to a washer to improve the grip with the machine clamps during the traction test.

\begin{figure}[!ht]
\centering
\includegraphics[width=3.4in]{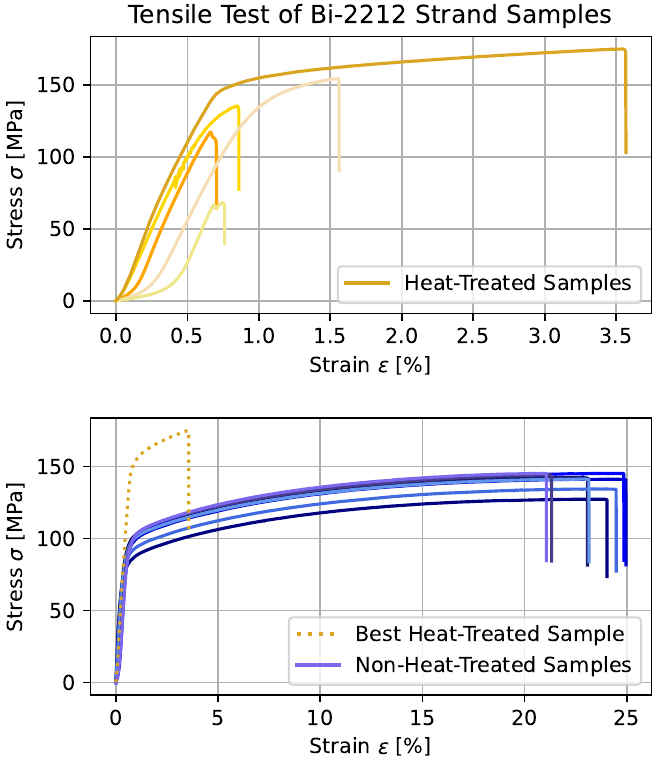}
\caption{Stress-strain curves of heat-treated (Top) and non-heat-treated (Bottom) samples of Bi-2212 strand tested with an Instron machine.}
\label{fig:SS_re}
\end{figure}

The goal of the tensile analysis was to obtain the conductor's Young modulus (E) along its axial direction. The analysis showed that the non-heat-treated and the heat-treated specimens have almost identical Young moduli, E$_{nHT}$= 21.99 $\pm$ 5.3\% GPa and E$_{HT}$= 21.26 $\pm$ 10.8\% GPa, respectively.
The two samples showed different yield strength results. To define the material properties of the Bi-2212 elements in the mechanical analysis, we used the measured E$_{HT}$ and the yield strength of the best heat-treated sample of approximately 140 MPa. However, the reacted strands broke unevenly during the plastic transition, suggesting that more data would be necessary to determine a more accurate yield strength of the Bi-2212 strand after the thermal reaction.
We needed to use the reacted samples data since, at the beginning of the simulation, the entire coil structure is intended to have already been appropriately heat-treated in a 50 bar oxygen flow and respecting the specific temperature profile \cite{ref5}.

\begin{table}
\centering
\caption{Material Properties at 300/4.2 K\label{tab:mats}}
\begin{tabular}{l c c c}  
\Xhline{4\arrayrulewidth}
\multirow{4}{*}{\thead{\textbf{} \\ \textbf{}}} & \multicolumn{3}{c}{\thead{\textbf{Properties value}}} \\
\cline{2-4}
                                                            &
\thead{\textbf{Young} \\ \textbf{Modulus} \\ \textbf{[GPa]}} &
\thead{\textbf{Yield} \\ \textbf{Strength} \\ \textbf{[MPa]}} &
\thead{\textbf{Thermal} \\ \textbf{Contraction} \\ \textbf{$\alpha \cdot 10^{-6} [K^{-1}]$}} \\
\hline
\thead[l]{Mullite\&CTD101K} & 12.9/19.7 & 790/1360& 26 \\ 
\thead[l]{Bi-2212} & 21.3 & 140 & 11.1\\ 
\thead[l]{Silver} & 82.5/91 & 54/80 & 14.6\\ 
\thead[l]{Inconel-718} & 208 & 1100/1700 & 8.2\\ 
\thead[l]{Iron} & 205 & 100/300 & 7\\ 
\thead[l]{Kapton} & 2.5 & 89.6 & 70\\ 
\thead[l]{SS 304} & 199/210 & 215/439 & 10.3\\ 
\Xhline{4\arrayrulewidth}
\end{tabular}
\end{table}

We referred to the literature for the mechanical properties of the structural materials at warm and cold temperatures, as reported in Table \ref{tab:mats}
\cite{ref3,ref10,ref11,ref12,ref13}.
In some cases (Inconel-718, iron, Kapton), the Young modulus variation between room and cold temperature is negligible, and we simply used one single value independent of temperature. This reduced the computational load of the analysis without significantly compromising the simulation accuracy.

\subsection{Mechanical Analysis}

The main goal of the mechanical FEM simulation is to compute the stress and strain distributions within the Bi-2212 and silver in the strands and compare the results obtained with the homogeneous and heterogeneous models. The mandrel structure of the two-layer coil doesn't change between the two simulations (Fig. \ref{fig:mech_struct}). Since the mandrel pieces are free to slide axially during heat treatment, and the grooves' design allows the insulated cable to move, we can neglect the residual stress state due to thermal cool-down after the reaction process. For this reason, the non-linear mechanical analysis follows a load-step approach which consists of three steps: a) pre-stress load, implemented as a 100 $\mu$m interference compression at the contact surface between the second layer skin and the iron yoke. That value was chosen to minimize coil displacements due to Lorentz forces. b) thermal cool-down from 300 K to 4.2 K, and c) energization at 10 kA, reaching 4.8 T and generating the computed Lorentz forces.
We used the element type \textsc{Plane 183}, a high-order 2D, 8-node element with quadratic displacement behavior for plasticity implementation (as for Bi-2212 in Fig. \ref{fig:SS_re}) and generalized plane strain options needed for mechanical simulation.
The magnet symmetry allows to model only one quarter of the structure cross-section with symmetry boundary condition on X and Y axes. The area elements follow a bi-linear perfectly plastic rheological behavior. The homogeneous model is characterized by the Bi-2212 anisotropic material properties reported in \cite{ref13}. The heterogeneous model is characterized by the isotropic material properties of Mullite plus epoxy, silver and Bi-2212 (Fig. \ref{fig:mech_struct}). We defined the conductors, mullite/epoxy and insulation areas with a finer mesh compared to the mandrel and iron yoke areas. We applied \textsc{Contact 172} and \textsc{Target 169} to the material interfaces of mandrels-epoxy, epoxy-ground insulation, and ground insulation-skin, with interference only along the contact highlighted by the green dashed arc in Fig. \ref{fig:mech_struct}.

\begin{figure}[!ht]
\centering
\includegraphics[width=3.4in]{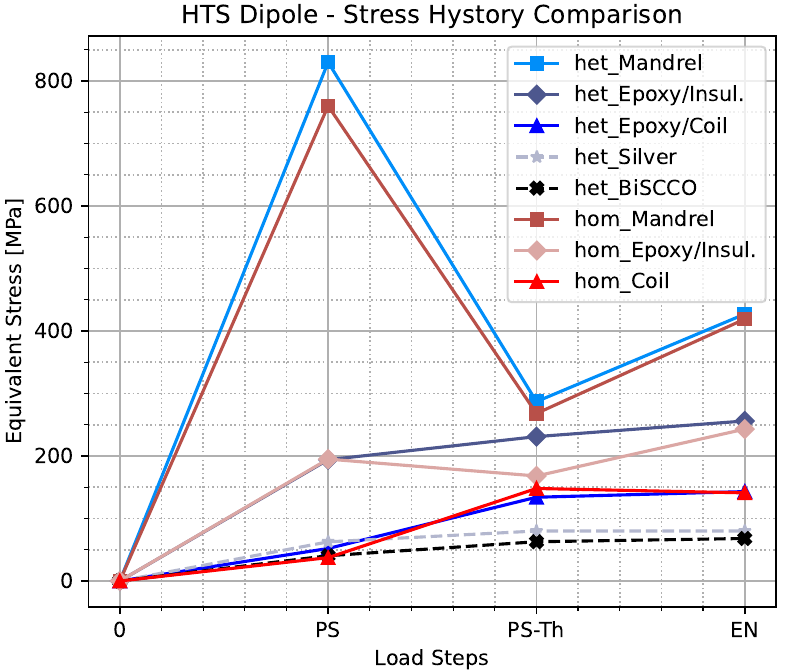}
\caption{Maximum equivalent Von Mises stress (MPa) for the insert components and the three load steps in the homogeneous and heterogeneous models.
The '\texttt{hom\_}' and  '\texttt{het\_}' refer to the homogeneous and heterogeneous model, respectively.}
\label{fig:StrHy}
\end{figure}

\begin{figure}[!ht]
\centering
 \includegraphics[width=3.2in]{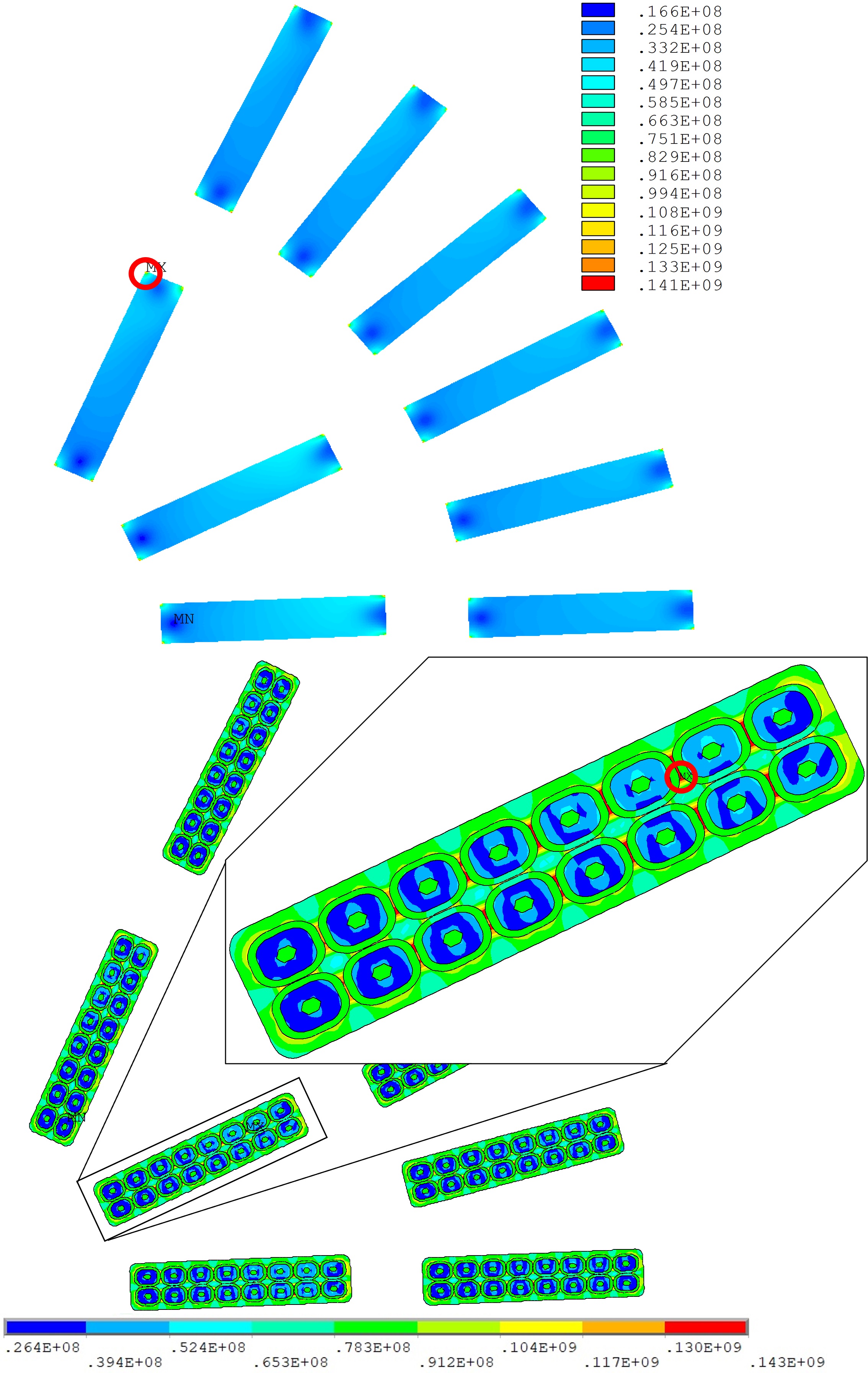}
\caption{Equivalent Von Mises stress distribution at EN load step (MPa) in the homogeneous (Top) and heterogeneous models (Bottom). The maximum stress is highlighted with a red circle in both of the models.}
\label{fig:mech_res_strcoil}
\end{figure}

Fig. \ref{fig:StrHy} shows the maximum equivalent Von Mises stress for each component of the insert (mandrel, epoxy insulation, epoxy coil, Bi-2212, silver) at each of the three simulation load steps (pre-stress 'PS', cool-down 'PS-Th', and energization 'EN' at 4.8 T) as computed in the two different simulations. Here are the results for each structural component.

\begin{itemize}
\item{Mandrel:
In the homogeneous (heterogeneous) model the maximum stress at energization (EN) is 419 (427) MPa at the bottom of the fourth groove from the mid-plane of the outer layer. Stress intensity in the mandrel drops at the cool-down load step since Inconel-718 is characterized by a higher thermal contraction factor compared to iron. For this reason, the pre-stress compression is partially relieved. The largest difference between the two models is 70 MPa at the bottom of the third groove from the mid-plane of the internal layer at pre-stress (PS). This is probably due to the different behavior of the heterogeneous and homogeneous cables inside the structure during compression.}
\item{Epoxy insulation:
In the homogeneous (heterogeneous) model the maximum stress at energization (EN) is 243 (256) MPa. The largest difference between the two models is 53 MPa at cool-down (PS-Th). We believe that this difference is due to the presence of fillets inserted between the mandrel and the epoxy insulation only in the heterogeneous model (Fig. \ref{fig:mech_struct}).}
\item{Conductor: In the homogeneous model, the maximum stress is 141 MPa at the edge of the cable. If only the area where the Bi-2212 is located in the heterogeneous model is considered, the maximum stress is 50 MPa. In the heterogeneous model, the maximum stress is 143 MPa in the epoxy coil, and 68 MPa in the Bi-2212 (Fig. \ref{fig:mech_res_strcoil}).
Silver reaches plasticity and the stress settles at its yield strength of 80 MPa. No other material reaches plasticity during the simulation load steps. In practice, homogeneous cable and the coil epoxy stress histories in the heterogeneous model are very similar.}
\end{itemize}

In Table \ref{tab:diffstress} are highlighted the differences between the equivalent Von Mises stress of components that could be directly compared between the two models (Mandrel, Epoxy insul. and Coil epoxy), at each load step.

\begin{table}[!ht]
\centering
\caption{Stress differences between the comparable components in the homogeneous and heterogeneous models [MPa] \label{tab:diffstress}}
\begin{tabular}{l c c c}  
\Xhline{3\arrayrulewidth}
\multirow{2}{*}{\thead{\textbf{}}} &       \multicolumn{3}{c}{\thead{\textbf{Components}}}     \\
\cline{2-4}
                                            & \thead{\textbf{Coil Ins.}} & \thead{\textbf{Mandrel}} & \thead{\textbf{Epoxy Ins.}}\\
\hline
\thead[l]{\textbf{PS}} & -14.6	& -70 &	1 \\
\thead[l]{\textbf{PS-Th}} & 14 & -19 &	-63 \\
\thead[l]{\textbf{EN}} & -2	& -8 &	-13 \\
\Xhline{3\arrayrulewidth}
\end{tabular}
\end{table}

Table \ref{tab:arstress} reports the maximum and minimum values of radial, azimuthal, and axial stress components for the homogeneous and heterogeneous models at the energization load step (EN). Directional stress, excluding the axial one, spans a $\sim$140 MPa range for the heterogeneous model, which is wider than the $\sim$120 MPa range estimated for the homogeneous model.
However, the maximum absolute value is lower for the heterogeneous model, where the radial and azimuthal stresses are 72 MPa in compression and 80 MPa in traction, respectively, compared to 118 MPa in compression and 108 MPa in compression for the homogeneous model.

\begin{table}[!ht]
\centering
\caption{Directional Stress Results in the cable [MPa] \label{tab:arstress}}
\begin{tabular}{p{3.5cm} c c}  
\Xhline{3\arrayrulewidth}
\thead{\textbf{Stress component (min/max)}} & \thead{\textbf{Homogeneous}} & \thead{\textbf{Heterogeneous}} \\
\hline
\thead{\textbf{Radial}} & -108/9 & -72/70 \\ 
\thead{\textbf{Azimuthal}} & -118/6 & -63/80 \\ 
\thead{\textbf{Axial}} & -30/6 & -21/101 \\ 
\Xhline{3\arrayrulewidth}
\end{tabular}
\end{table}

\begin{figure}[!ht]
\centering
\includegraphics[width=3in]{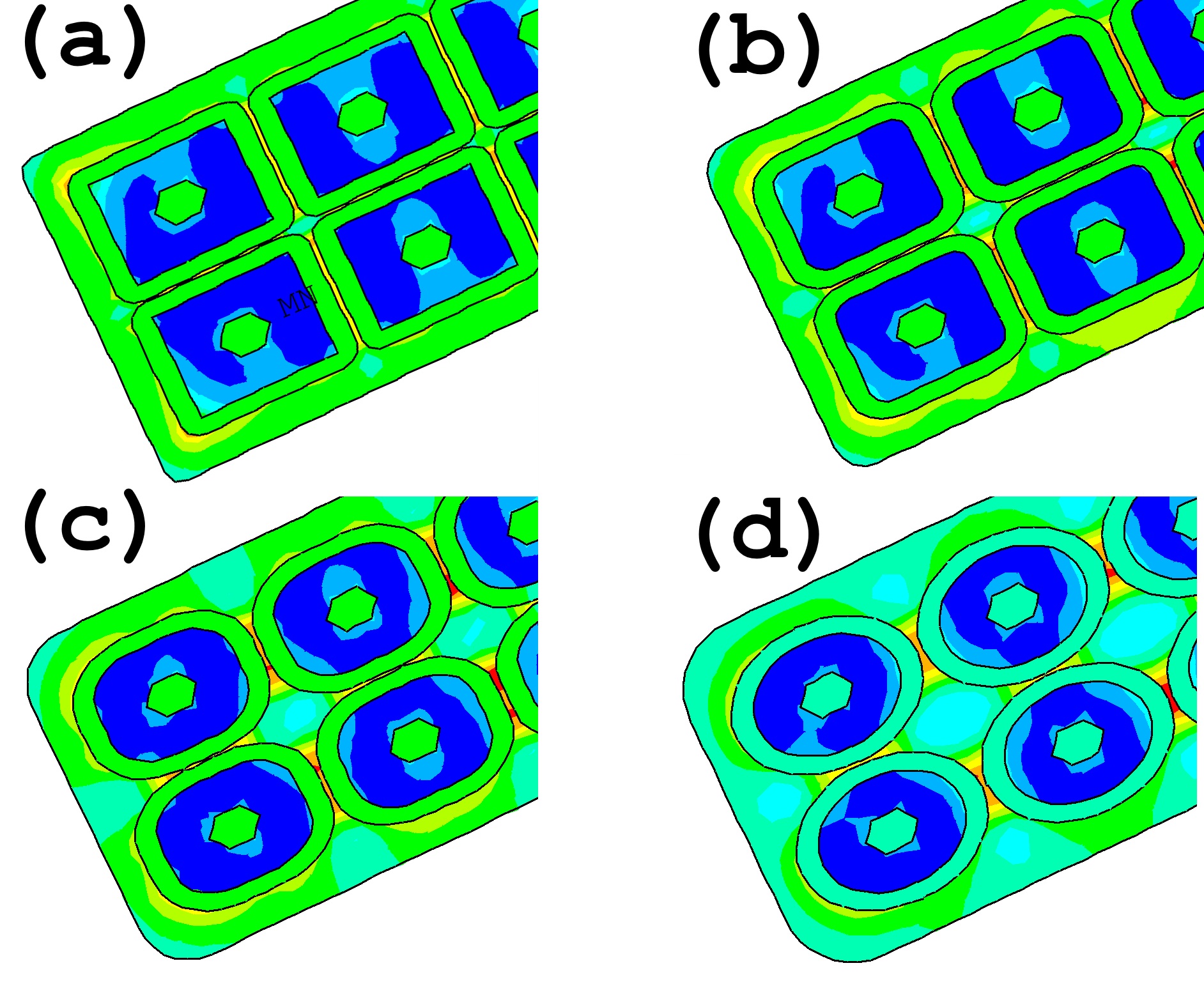}
\caption{Variation of the cable geometry and stress distribution between four different SP values: (a) 0.0, (b) 0.3, (c) 0.7, (d) 1.0.}
\label{fig:StrPar}
\end{figure}

We also performed a sensitivity analysis to search for possible correlations between stress and the geometrical deformation of the strand in the cable. The  Bi-2212 strands can be deformed during the cold-rolling extrusion of the cable from the Turks Head. To reproduce this effect, we developed an APDL macro to modify the cable with a strand shape parameter (SP). If SP = 1, strands have an elliptical shape. If SP = 0, strands are completely packed and compressed inside the cable area (Fig. \ref{fig:StrPar}). Table \ref{tab:sens} reports the maximum equivalent stress in all the different structural components at the energization load step (EN) for four representative SP values (SP = 0.0, 0.3, 0.7, and 1.0).
Table \ref{tab:sens} shows that the configuration with SP = 0.7 is the least conservative one concerning the stress value inside the Bi-2212 since the stress reaches its minimum value.
However, the structural materials show moderate sensitivity to the stress as SP varies. 
Although this is the least conservative configuration, we chose SP = 0.7 since it better matches the actual Bi-2212 strand geometry. 

\begin{table}[!ht]
\centering
\caption{Von Mises Stress dependence on the Strand Shape Parameter (max/min) [MPa] \label{tab:sens}}
\begin{tabular}{l c c c c}  
\Xhline{3\arrayrulewidth}
\multirow{2}{*}{\thead{\textbf{}}} &       \multicolumn{3}{c}{\thead{\textbf{Strand Shape Parameter (SP) }}}     \\
\cline{2-5}
                                            & \thead{\textbf{0.0}} & \thead{\textbf{0.3}} & \thead{\textbf{0.7}} & \thead{\textbf{1.0}} \\
\hline
\thead[l]{\textbf{Bi-2212}} & 96/25 & 73/26 & 68/26 & 78/25 \\
\thead[l]{\textbf{Silver}} & 80/63 & 80/68 & 80/67 & 80/61 \\
\thead[l]{\textbf{Coil Epoxy}} & 138/61 & 137/59 & 143/60 & 157/62 \\
\thead[l]{\textbf{Epoxy insulation}} & 250/69 & 254/69 & 256/69 & 259/69 \\
\thead[l]{\textbf{Mandrel}} & 422/0.3 & 423/0.5 & 427/0.5 & 429/0.2 \\
\Xhline{3\arrayrulewidth}
\end{tabular}
\end{table}

\section{Conclusion}
In this paper, the up-to-date design of the Bi-2212 insert coil developed at Fermilab was presented. 
We reported the results of the magnetic and mechanical simulations performed with ROXIE and ANSYS. We developed two alternative models of the Bi-2212 cable. In the "homogeneous model", the cable is represented in the simulation as a simple rectangle of homogeneous material. In the "heterogeneous model", all cable components are modeled at the strand level (Sect. I).
The magnetic analysis shows no significant differences in the magnetic field distribution and Lorentz force intensity between the two models (Sect. III.A).
To obtain the Bi-2212 cable mechanical properties needed for the mechanical simulation of the heterogeneous model, we performed a series of measurements with a tensile test Instron machine at Fermilab to plot the Bi-2212 strands stress-strain curves (Sect. III.B). Those measurements allow us to set the heat-treated superconductor Young modulus value at 21.3 GPa and the yield stress at 140 MPa in the mechanical analysis.
We computed the maximum equivalent Von Mises stress for the three load steps (pre-stress (PS), cool-down (PS-Th), and energization (EN) at 4.8 T) in the insert components for both the homogeneous and the heterogeneous model (Sec. III.C).
For the homogeneous model, the maximum stress in the conductor is 141 MPa located on the edge of the cable. That value can be reduced by optimizing pre-compression load and shim thickness to reduce stresses in the coil instead of minimizing coil displacement after energization. Considering, instead, the area within the homogeneous Rutherford cable where the Bi-2212 strands should be located, the maximum stress is 50 MPa. 
With the heterogeneous model, the maximum stress in the Bi-2212 areas is 68 MPa, which respects the threshold of 120 MPa \cite{ref13}. The maximum stress in the structural components remains within the acceptable value of yield strength set for their material, except for silver and iron yoke where plasticity occurs. The accuracy of the Bi-2212 dipole insert mechanical analysis was improved thanks to the heterogeneous model, which gives more detailed information about stress state distribution and intensity in all the coil components. 
We also investigated the dependence of the equivalent Von Mises stress in the cable on the strand deformation (SP parameter). The higher sensitivity is in the Bi-2212 conductor area since silver reaches plasticity, and the strand shape variation affects only indirectly the other structural materials. Even if this is the least conservative configuration, we used the SP value to 0.7 when comparing the heterogeneous and homogeneous models, since it represents the most realistic cable geometry. That SP value will be set as standard for the future FEM analysis based on the heterogeneous Rutherford cable model.

\section*{Acknowledgments}
This work was supported by the EU Horizon 2020 Research
and Innovation Program under the Marie Sklodowska-Curie Grant Agreement No. 734303, 822185, 858199 and 101003460, and the Horizon Europe Research and Innovation Program under the Marie Sklodowska-Curie Grant Agreement No. 101081478. We want to thank collaborators from CEA Paris-Saclay, Lawrence Berkeley National Laboratory, and CERN for their technical advice and suggestions, which enhanced the quality of the work published in this article.

\vfill

\end{document}